# ISDNN: A DEEP NEURAL NETWORK FOR CHANNEL ESTIMATION IN MASSIVE MIMO SYSTEMS

ISDNN: MẠNG NƠ-RON SÂU CHO ƯỚC LƯỢNG KÊNH HỆ THỐNG MASSIVE MIMO

*Do Hai Son[1], Vu Tung Lam[2], Tran Thi Thuy Quynh[2,*]*

[1]*Information Technology Institute, Vietnam National University, Hanoi*

[2]*University of Engineering and Technology, Vietnam National University, Hanoi*

[*]*Email: quynhttt@vnu.edu.vn*

**ABSTRACT**

Massive Multiple-Input Multiple-Output (massive MIMO) technology stands as a cornerstone in 5G and beyonds. Despite the remarkable advancements offered by massive MIMO technology, the extreme number of antennas introduces challenges during the channel estimation (CE) phase. In this paper, we propose a single-step Deep Neural Network (DNN) for CE, termed Iterative Sequential DNN (ISDNN), inspired by recent developments in data detection algorithms. ISDNN is a DNN based on the projected gradient descent algorithm for CE problems, with the iterative iterations transforming into a DNN using the deep unfolding method. Furthermore, we introduce the structured channel ISDNN (S-ISDNN), extending ISDNN to incorporate side information such as directions of signals and antenna array configurations for enhanced CE. Simulation results highlight that ISDNN significantly outperforms another DNN-based CE (DetNet), in terms of training time (13%), running time (4.6%), and accuracy (0.43 dB). Furthermore, the S-ISDNN demonstrates even faster than ISDNN in terms of training time, though its overall performance still requires further improvement.

**TÓM TẮT**

Massive MIMO là một công nghệ nền tảng được sử dụng trong các hệ thống truyền thông 5G trở lên. Mặc dù mang lại nhiều lợi thế nhưng công nghệ này cũng gặp thách thức lớn về độ phức tạp tính toán trong pha ước lượng kênh do số lượng rất lớn các phần tử anten trong mảng. Bài báo này đề xuất một mạng nơ-ron sâu đơn bước mở, được đặt tên là ISDNN (Iterative Sequential Deep Neural Network) nhằm cải thiện độ phức tạp tính toán trong ước lượng kênh massive MIMO. Ý tưởng xây dựng ISDNN là áp dụng kỹ thuật trải sâu cho một giải thuật lặp để ước lượng kênh, mỗi lớp trong mạng thực thi một lần lặp, các thông số vào ban đầu được tính dựa trên thuật toán ước lượng phổ biến LS (Least Square). Hơn nữa, bài báo cũng thực hiện việc mở rộng ISDNN thành S-ISDNN (structured channel ISDNN) để áp dụng cho trường hợp kênh có cấu trúc. Kết quả nghiên cứu chỉ ra rằng, việc sử dụng ISDNN vượt trội khi so sánh với một mô hình mạng đã được đề xuất trước đây là DetNet, về thời gian đào tạo (13%), thời gian chạy (4.6%), và độ chính xác (tốt hơn 0.43 dB). Hơn nữa, S-ISDNN còn có thời gian đào tạo nhanh hơn so với ISDNN, mặc dù hiệu năng tổng thể của nó vẫn cần được cải thiện thêm.

***Từ khóa:*** *MIMO siêu lớn, ước lượng kênh, mạng nơ-ron sâu một bước, mô hình kênh không sử dụng cấu trúc/có cấu trúc.*

## 1. INTRODUCTION

Massive MIMO is an essential technology in 5G and beyonds. This technology offers significant improvements in spectral efficiency and capacity by exploiting a large number of antennas to serve multiple users simultaneously. Moreover, with the base station (BS) employing thousands of antennas, the wireless channel experiences channel hardening characterized by predominant large-scale fading effects and extended coherence time [1]. However, the proliferation of antennas also leads to more complexity during the CE phase due to the intricate structure of the channel matrix, which may not always exhibit full rank. Thus, low-complexity CE algorithms have attracted a lot of studies.

In this paper, we focus on applying deep learning (DL), a trending approach, in CE for massive MIMO systems [2]. In one of the first studies [3], the authors employed a convolutional neural network (CNN) to estimate the channel in millimeter wave (mmWave) massive MIMO systems. This work used combining matrix, beamforming matrix, and assuming pilot sequences are all a constant value to pre-estimate a "tentatively estimated channel". The CNN network corresponds to denoising a quasi-accurate channel. In [4], the authors proposed to use an Auto-Encoder network for CE. In this study, pilot sequences are randomly generated. Therefore, at the first step of CE, [4] remove known pilots from received signals using an inverse transformation (Least-square method). Likewise, [5] used a deep neural network (DNN) to first denoise the received signal, followed by a conventional least-squares (LS) estimation. In these studies, a common aspect is the requirement for two steps for CE, such as (i) removing the known pilot signal from the received signal and (ii) the precise estimation of the channel.

Inspired by data detection algorithms in [6], [7], we propose a single-step DNN for CE, namely ISDNN (Iterative Sequential Deep Neural Network). We convert the iterative sequential algorithm in [7] to a DNN by the deep unfolding method [8] and modify the structure of DNN to achieve better accuracy. In addition, the complexity of ISDNN is reduced since it does not require the inversion/pseudo-inversion operation as an LS estimator. Instead, ISDNN approximates this operation through the learning process. This paper also extends ISDNN to S-ISDNN (structured channel ISDNN) for estimating mm-wave channels with known directions of arrival and array geometry (side information). The mm-wave channel is usually applied for next-generation wireless communications and named as "structured channel" with few dominant rays, also used in our previous research [9]. According to simulation results, the proposed ISDNN structure outperforms DetNet-based CE [6] in terms of training time, running time, and accuracy. Moreover, the S-ISDNN demonstrates even faster than ISDNN regarding training time, but its performance requires further improvement.

The main contribution of this paper is to propose ISDNN and S-ISDNN estimators for CE and side information-aided CE, respectively. The massive MIMO system model is presented in section 2. The proposed ISDNN and S-ISDNN are shown in section 3. Finally, we conduct simulations to compare the performance of ISDNN with that of another DNN-based estimator. The source code is available at [1].

## 2. SYSTEM MODEL

In this work, we consider a massive MIMO system, which consists of $N_t$ transmit antennas and $N_r$ receive antennas ($N_t \ll N_r$). In the up-link channel, at time $n$, the Eq. (1) expresses the system model.

$$\mathbf{y}(n) = \mathbf{x}(n)\mathbf{H}(n) + \mathbf{w}(n), \qquad (1)$$

---

[1] https://github.com/DoHaiSon/ISDNN

where $\mathbf{x} \in \mathbb{C}^{1 \times N_t}$, $\mathbf{y} \in \mathbb{C}^{1 \times N_r}$, and $\mathbf{w} \in \mathbb{C}^{1 \times N_r}$ are vectors of transmit signal, receive signal, and additive noise, respectively. We assume that the elements of $\mathbf{w}(n)$ are independent and identically distributed (i.i.d) random variables following a complex normal distribution $\mathcal{CN}(0, \sigma^2 I)$. The effects of wireless propagation are represented by $\mathbf{H} \in \mathbb{C}^{N_t \times N_r}$, a.k.a channel matrix. Hereafter, for brevity, we omit the timestamp $n$. Note that the condition $N_t \ll N_r$ is crucial. In the contrary scenario, the CE problem, given $\mathbf{x}$ and $\mathbf{y}$, becomes underdetermined. The elements in $\mathbf{H}$ are represented as complex numbers, capturing both the amplitude and phase effects induced by the channel. Without loss of generality, complex values are often decomposed into real ($\mathfrak{R}$) and imaginary ($\mathfrak{J}$) parts, as follows:

$$\mathbf{x} = [\mathfrak{R}(\mathbf{x}) \quad \mathfrak{J}(\mathbf{x})]; \qquad \mathbf{y} = [\mathfrak{R}(\mathbf{y}) \quad \mathfrak{J}(\mathbf{y})];$$
$$\mathbf{w} = [\mathfrak{R}(\mathbf{w}) \quad \mathfrak{J}(\mathbf{w})]; \quad \mathbf{H} = \begin{bmatrix} \mathfrak{R}(\mathbf{H}) & -\mathfrak{J}(\mathbf{H}) \\ \mathfrak{J}(\mathbf{H}) & \mathfrak{R}(\mathbf{H}) \end{bmatrix}^\top, \tag{2}$$

with $\mathbf{x} \in \mathbb{R}^{1 \times 2N_r}$, $\mathbf{y} \in \mathbb{R}^{1 \times 2N_r}$, $\mathbf{w} \in \mathbb{R}^{1 \times 2N_r}$, and $\mathbf{H} \in \mathbb{R}^{2N_t \times 2N_r}$. The $(.)^\top$ is the transpose operator.

In the CE process, pilot sequences are organized into three types: block-type, comb-type, and lattice-type [10]. This study adopts the block-type arrangement, where all channels transmit pilot signals at specific time slots. This implies a prior knowledge of $x$ and $y$ in Eq. (1) for CE. Subsequently, the estimated channel is utilized in multiple subsequent data time slots within a channel coherence time [11].

In the estimation process, $\mathcal{L}\left(\mathbf{H}; \widehat{\mathbf{H}}_\Theta(\mathbf{x}, \mathbf{y})\right)$ is the lost function between the original $\mathbf{H}$ and estimated $\widehat{\mathbf{H}}$ given $\mathbf{x}$, $\mathbf{y}$. By using the Eq. (3), we can find the best channel matrix.

$$\min_{\Theta} \mathbb{E}\{\mathcal{L}(\mathbf{H}; \widehat{\mathbf{H}}_\Theta(\mathbf{x}, \mathbf{y}))\}, \tag{3}$$

where $\mathbb{E}$ and $\Theta$ are the expectation value and learning values, respectively.

## 3. THE PROPOSED ISDNN

### 3.1. The architecture of the proposed ISDNN

The optimal solution to solve Eq. (3) is the Maximum-likelihood estimator (MLE), as follows:

$$\widehat{\mathbf{H}}_\Theta(x, y) = \underset{\mathbf{H} \in \mathbb{R}^{2N_t \times 2N_r}}{\arg\min} \; |\mathbf{y} - \mathbf{x}\mathbf{H}|^2, \tag{4}$$

However, the computational complexity of MLE increases exponentially with $N_t$ and $N_r$. Thus, it is unfeasible to deploy it in the massive MIMO system of interest. Inspired by [6], the projected gradient descent (PGD) method and chain rule are used to solve the Eq. (4). The solution is given by Eq. (5).

$$\widehat{\mathbf{H}}_{k+1} = \mathbf{\Gamma}\left[\widehat{\mathbf{H}}_k - \delta_k \frac{\partial \parallel \mathbf{y} - \mathbf{x}\mathbf{H} \parallel^2}{\partial \mathbf{x}}\bigg|_{\mathbf{H}=\widehat{\mathbf{H}}_k}\right]$$
$$= \mathbf{\Gamma}[\widehat{\mathbf{H}}_k - \delta_k \mathbf{x}^\top \mathbf{y} + \delta_k \mathbf{x}^\top \mathbf{x} \widehat{\mathbf{H}}_k], \tag{5}$$

where $\widehat{\mathbf{H}}_k$ is the estimated channel matrix at the $k$-th iteration, for $k = 1, \dots, K$. $\mathbf{\Gamma}$ is a non-linear operator and $\delta_k$ is the learning rate. In [6], the authors unfolded this iterative solution to a DNN structure, named DetNet, for data detection problems. In this paper, we propose a new structure of the network for channel estimation, called ISDNN.

In [6], the authors initialized $\widehat{\mathbf{H}}_1$ by a zeros matrix. However, [13] pointed out that by taking advantage of the LS estimator, the learning process can be accelerated. The formula of LS-based estimator is expressed by Eq. (6).

$$\widehat{\mathbf{H}}_{\text{LS}} = (\mathbf{x}^\top \mathbf{x})^{-1} \mathbf{x}^\top \mathbf{y} = \mathbf{P}^{-1} \mathbf{q}. \tag{6}$$

Note that due to the transformation in Eq. (2), the Hermitian operator $(.)^H$ is turned into transpose. We denote $\mathbf{G}_\mathbf{x} = \mathbf{x}^\top \mathbf{x}$ and $\mathbf{q} = \mathbf{x}^\top \mathbf{y}$. The diagonal elements of $\mathbf{G}_\mathbf{x}$ are combined into a diagonal matrix $\mathbf{D} = \text{diag}(\mathbf{G}_\mathbf{x})$. At the initialization step, the initial $\mathbf{H}_1$ to be input to the first layer of ISDNN is given by Eq. (7).

$$\mathbf{H}_1 = \mathbf{D}^{-1} \mathbf{q}. \tag{7}$$

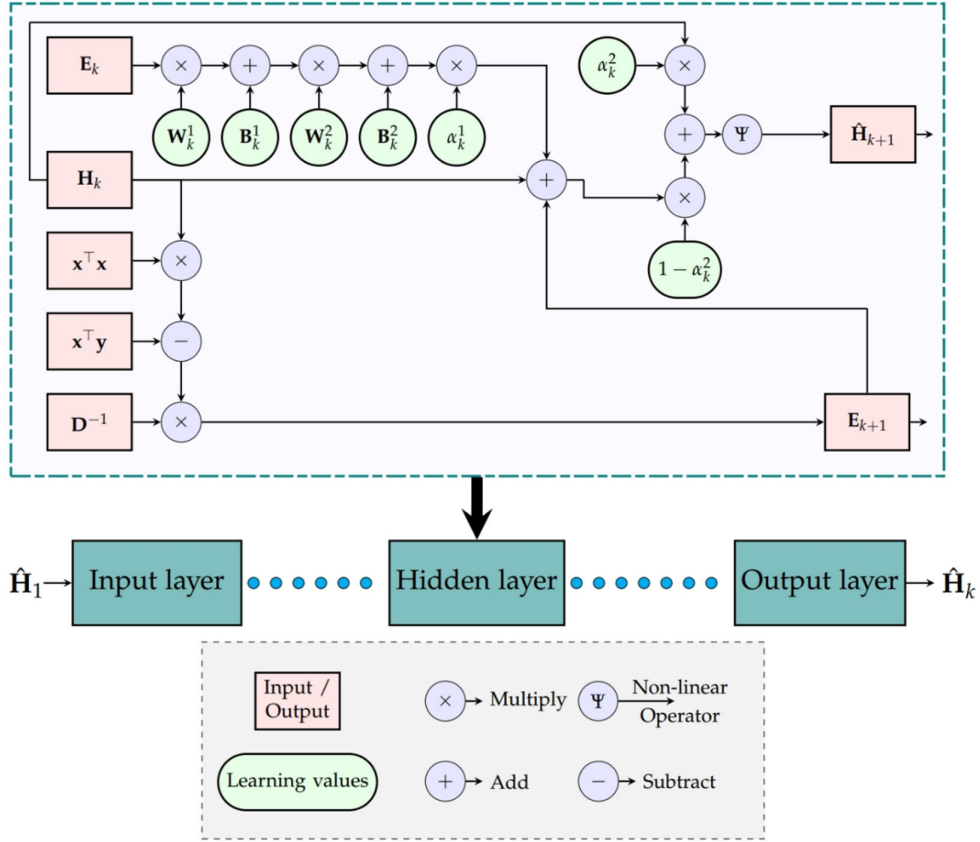

*Figure 1: The architecture of a layer in the proposed ISDNN network*

Figure 1 shows the architecture of a layer in our proposed ISDNN network. From the ISDNN's second layer onwards, $\mathbf{E}_k^j$ is residual matrix at $j$-th iteration and $k$-th layer can be computed by Eq. (8).

$$\mathbf{E}_k^j = \mathbf{y} - \mathbf{x} \mathbf{H}_k^j, \tag{8}$$

Note that, in [7], the authors proved that $|\mathbf{E}_k^j|^2 < |\mathbf{E}_k^{j-1}|^2$ for the data detector problems. Hence, the distance between $\mathbf{H}$ and $\widehat{\mathbf{H}}$ is reduced after iterations. After any layer, ISDNN [7] updates the estimated $\widehat{\mathbf{H}}$ by Eq. (9).

$$\widehat{\mathbf{H}}_{k+1} = \widehat{\mathbf{H}}_k + \mathbf{E}_{k+1}, \tag{9}$$

The $\mathbf{E}_{k+1}$ in Eq. (9) can be re-write as the decomposed approach: $x^\top y = x^\top x H + x^\top w$, as follows:

$$\mathbf{E}_{k+1} = \mathbf{D}^{-1}(\mathbf{x}^\top \mathbf{y} - \mathbf{x}^\top \mathbf{x} \widehat{\mathbf{H}}_k). \tag{10}$$

However, $\widehat{\mathbf{H}}_{k+1}$ depends on not only $\mathbf{E}_{k+1}$ as shown in Eq. (9) but also residual matrices from previous layers, i.e., $\mathbf{E}_k$, $\mathbf{E}_{k-1}$,…, $\mathbf{E}_1$. Nevertheless, due to the highest correlation between neighboring residual matrices, the ISDNN only considers the influence of $\mathbf{E}_k$ at the $k$-th layer. Thus, we add a learning value $\alpha_k^1$ to each ISDNN layer to show residual matrices' impacts.

$$\boldsymbol{\mu}_k = \widehat{\mathbf{H}}_k + \mathbf{E}_{k+1} + \alpha_k^1 \mathbf{E}_k. \tag{11}$$

Moreover, we do not directly assign $\widehat{\mathbf{H}}_{k+1} = \boldsymbol{\mu}_k$. We use the convex combination [12] of $\boldsymbol{\mu}$ and $\widehat{\mathbf{H}}$ with learning values $\alpha_k^2$ to consider the association between them. In this way, $\widehat{\mathbf{H}}_{k+1}$ is contributed by both $\boldsymbol{\mu}_k$ and $\widehat{\mathbf{H}}_k$ in $\alpha_k^2$ ratio. Eq. (9) is turned into Eq. (12).

$$\widehat{\mathbf{H}}_{k+1} = (1 - \alpha_k^2)\boldsymbol{\mu}_k + \alpha_k^2 \widehat{\mathbf{H}}_k. \tag{12}$$

After that, we add two linear operators before updating $\mathbf{E}_k$. This modification increases training time but will also increase accuracy in the case of complex channels.

$$\mathbf{E}_k \leftarrow \mathbf{W}_k^2(\mathbf{W}_k^1 \mathbf{E}_k + \mathbf{B}_k^1) + \mathbf{B}_k^2. \tag{13}$$

where $\mathbf{W}_k$ and $\mathbf{B}_k$ are matrices of weight and bias in a *Pytorch* framework linear function.

The detectors in [6, 7, 13] sequentially feed the training data corresponding to a single sample into the network. This significantly reduces the learning speed of a DNN network by not leveraging the advantages of tensor data types, which allow the manipulation of multi-dimensional variables. Therefore, the next improvement point of the ISDNN is to use a much larger 'Batch size' ($bs$), where '$bs$' represents the amount of data used in a single iteration. The input data of the ISDNN are combined into a 3-dimensional tensor, as follows:

$$\widetilde{\mathbf{E}}_k \leftarrow [\mathbf{E}_k^{j,1}, \mathbf{E}_k^{j,2}, \dots, \mathbf{E}_k^{j,bs}]; \quad \widetilde{\mathbf{H}}_k \leftarrow [\widehat{\mathbf{H}}_k^{j,1}, \widehat{\mathbf{H}}_k^{j,2}, \dots, \widehat{\mathbf{H}}_k^{j,bs}];$$
$$\widetilde{\mathbf{X}} \leftarrow [\mathbf{x}^{j,1}, \mathbf{x}^{j,2}, \dots, \mathbf{x}^{j,bs}]; \qquad \widetilde{\mathbf{y}} \leftarrow [\mathbf{y}^{j,1}, \mathbf{y}^{j,2}, \dots, \mathbf{y}^{j,bs}].$$

However, this is a technical programming proposal, mathematical symbols such as matrix multiplication notation similar to $bs = 1$ will still be retained to avoid confusion.

### 3.2. Learning process of ISDNN

The learning values of the training process are as follows: $\boldsymbol{\Theta} = \{\mathbf{W}_k^1, \mathbf{B}_k^1, \mathbf{W}_k^2, \mathbf{B}_k^2, \alpha_k^1, \alpha_k^2\}|_{k=1}^K$. The normalized mean square error (NMSE) function is utilized, as in Eq. (14), to measure the accuracy of the ISDNN network.

$$\text{NMSE}\left(\mathbf{H}; \widehat{\mathbf{H}}_{\boldsymbol{\Theta}}(\mathbf{x}, \mathbf{y})\right) = \frac{\sum_{t=1}^{N_t} \sum_{l=1}^{N_r} |h_{t,l} - \hat{h}_{t,l}|_F^2}{\sum_{t=1}^{N_t} \sum_{l=1}^{N_r} |h_{t,l}|^2}, \tag{14}$$

where $t, l$ are $t$-th transmitter and $l$-th receiver, respectively. $h_{t,l}$ is the element of row $t$-th, column $l$-th in the channel matrices $\mathbf{H}$ and $\widehat{\mathbf{H}}$. The four steps of an iteration in the training phase are as follows:

1) Initialize the initial parameters and residual vectors of the ISDNN structure: $\mathbf{H}_1, \mathbf{E}_1, \alpha_1^1, \alpha_1^2$.
2) Feed the dataset through the $K$ layers (forward propagation), then calculate the loss through the function built-in MSE function of *Pytorch* framework, as Eq. (15).

$$\mathcal{L}\left(\mathbf{H}; \widehat{\mathbf{H}}_{\boldsymbol{\Theta}}(\mathbf{x}, \mathbf{y})\right) = \frac{1}{N_t N_r} \sum_{t=1}^{N_t} \sum_{l=1}^{N_r} \left(h_{t,l} - \hat{h}_{t,l}\right)^2. \tag{15}$$

3) Back-propagate $\mathcal{L}\left(\mathbf{H}; \widehat{\mathbf{H}}_{\boldsymbol{\Theta}}(\mathbf{x}, \mathbf{y})\right)$ to obtain the gradient.
4) From the obtained gradient, ISDNN uses an optimization algorithm, such as Adam [14], to update the learning values $\boldsymbol{\Theta}$.

### 3.3. Structured channel ISDNN

In [9, 15], the authors presented channel models, i.e., "unstructured" and "structured", for massive MIMO and mmWave systems. In the unstructured channel model, propagation effects between transmit and receive antenna pairs are described by complex gains, while the structured model involves complex gains, DoA, and DoD. The elements in **H** are expressed as follows:

$$h_{t,l} = \sum_{p=0}^{P-1} \beta_{p,t} \cdot e^{-ik_s c_l(\theta_{p,t}, \phi_{p,t})}, \tag{16}$$

for the $p$-th ray, $\beta$ represents complex path gain, and "·" is the scalar product. Zenith and azimuth angles of DoA are $\theta, \phi$, respectively. Other notations are calculated [9] by $k_s = 2\pi/\lambda$; $c_l(\theta_{p,t}, \phi_{p,t}) = \hat{\boldsymbol{c}} \cdot \boldsymbol{c}_l$; $\hat{\boldsymbol{c}} = \sin\theta_{p,t}\cos\phi_{p,t}\hat{\boldsymbol{x}} + \sin\theta_{p,t}\sin\phi_{p,t}\hat{\boldsymbol{y}} + \cos\theta_{p,t}\hat{\boldsymbol{z}}$; $\boldsymbol{c}_l = x_l\hat{\boldsymbol{x}} + y_l\hat{\boldsymbol{y}} + z_l\hat{\boldsymbol{z}}$, where $\lambda$ is the wavelength; $\hat{\boldsymbol{c}}$ is the unit vector in the direction of the field point; $\boldsymbol{c}_l$ is the position of $l$-th element in receiver's antenna array $(x_l, y_l, z_l)$. In this work, we consider a simple case, which is the line-of-sight (LoS), i.e., $P = 1$, in terms of the structured channel model. Hence, Eq. (1) can be given by Eq. (17).

$$\mathbf{y} = \mathbf{xH} + \mathbf{w} = \mathbf{x}(\boldsymbol{\beta} \cdot \boldsymbol{\varphi}) + \mathbf{w} \tag{17}$$

where **β, φ** are matrices of $\beta_{p,t}$ and $e^{-ik_s c_l(\theta_{p,t}, \phi_{p,t})}$, thanks to the property of scalar product. In 5G and beyond wireless communication standards, DoA ($\boldsymbol{\theta}, \boldsymbol{\phi}$) of user equipments have been estimated and are available at the BS prior to CE. Therefore, DoA together with the configuration of the receive antenna array, is referred to as side information in this case. The ISDNN architecture will be modified to fit the structured channel model as described in Eq. (16) (referred to as "structured channel ISDNN", S-ISDNN). That leads to instead of estimating **H**, S-ISDNN will estimate **β**. Hence, Eq. (17) can be turned into Eq. (18).

$$\bar{\mathbf{y}} = \mathbf{x}\hat{\boldsymbol{\beta}} + \bar{\mathbf{w}} \quad \text{with} \quad \hat{\beta}_{t,l} = \frac{h_{t,l}}{\varphi_{t,l}} = \frac{h_{t,l}}{e^{-ik_s c_l(\theta_{1,t}, \phi_{1,t})}} \tag{18}$$

## 4. RESULTS AND DISCUSSION

In this section, we present the experimental analysis of the proposed ISDNN using the simulation parameters outlined in Table 1. Our experiments are run on a personal computer with processor Intel Core i9-10900 @5.2 GHz, RAM of 64GB, and GPU of RTX 3090 48GB VRAM. The ISDNN, S-ISDNN, and DetNet-based CE are programmed in Python and the well-known *PyTorch* framework. The training process is accelerated by GPU on a computer with no other processes running. The NMSE of trained ISDNN is the average of 100 testing times, as shown in Eq. (19), at each SNR level. The pilot sequences are assumed to modulate in 16-QAM type. Channel matrices **H** are generated as i.i.d Rayleigh channels [9] with $\mathcal{N}(0, 1/\sqrt{2})$.

$$\text{NMSE}\left(\mathbf{H}; \widehat{\mathbf{H}}_{\boldsymbol{\Theta}}(\mathbf{x}, \mathbf{y})\right) = \frac{\sum_{r=1}^{100} \text{NMSE}_r\left(\mathbf{H}; \widehat{\mathbf{H}}_{\boldsymbol{\Theta}}(\mathbf{x}, \mathbf{y})\right)}{100}. \tag{19}$$

*Table 1: Simulation parameters of the proposed ISDNN and wireless communications system*

| Parameters | Specifications | Parameters | Specifications |
|---|---|---|---|
| Massive MIMO system size | $N_t = 8, N_r = 64$ | Non-linear operator (**Ψ**) | Tanh |
| Modulation type | 16-QAM | Linear operator (**W**) size | $2N_r \times 2N_r$ |
| SNR levels of training dataset | [0, 5, 10, 15, 20] dB | $\alpha_k^1$ | $\mathcal{U}[0\ 1)$ |
| Training size | 50,000 samples | $\alpha_k^2$ | 0.5 |
| Testing size | 10,000 samples | $\mathbf{E}_1$ | $\mathcal{U}[0\ 1)$ |
| Optimization algorithm | Adam [14] | Learning rate | $\delta = 0.0001$ |

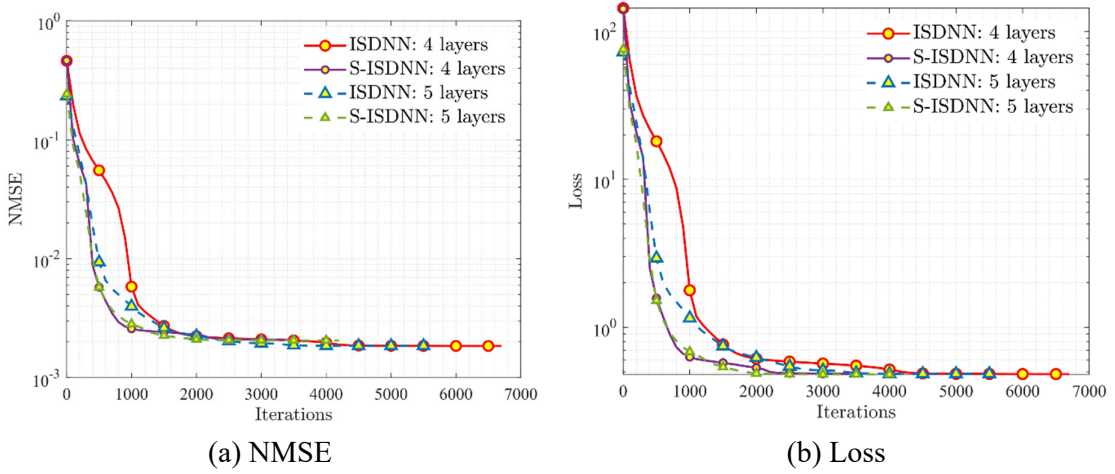

(a) NMSE  (b) Loss

*Figure 2: Training process of ISDNN and S-ISDNN*

Figure 2 illustrates the convergence of proposed ISDNN and S-ISDNN with the configuration of 4 and 5 layers. We do not train ISDNN by a fixed number of iterations but employ the "early stopping" function. As a result, if NMSE fails to decrease for 3 consecutive iterations, the training phase is terminated. In Figure 2a, in terms of accuracy, ISDNN with 4 layers is the best model with the latest NMSE ≈ 0.00184. The ISDNN with 5 layers ranks second with the latest NMSE ≈ 0.00185. In spite of side information-aided during the CE process, NMSEs of S-ISDNN networks are not as good as expected. The S-ISDNN with 4 and 5 layers have the latest NMSE values at 0.00209 and 0.00206, respectively. Regarding iteration, ISDNN with 4 and 5 layers and S-ISDNN with 4 and 5 layers, respectively, need 6700, 5600, 3400, and 4200 iterations to converge. Nonetheless, Table 2 reveals that fewer iterations do not necessarily equate to faster training time. Because the time per iteration of ISDNN, S-ISDNN with 4 layers is faster than that of ISDNN, S-ISDNN with 5 layers. The ISDNN and S-ISDNN consist of 4 layers, saving around 30% of training time compared to those with 5 layers. Additionally, despite its lower accuracy, S-ISDNN only needs half the number of iterations compared to ISDNN with 4 layers to converge. This can also be a trade-off during the training process. Figure 2b describes values of $\mathcal{L}\left(\mathbf{H}; \widehat{\mathbf{H}}_{\boldsymbol{\theta}}(\mathbf{x}, \mathbf{y})\right)$ computed by the MSE function. First, we can observe the same trend as NMSE in Figure 2a when $\mathcal{L}$ of ISDNN with 4 layers is the lowest. The latest loss values of ISDNN with 4, 5 layers and S-ISDNN with 4, 5 layers are 0.482667, 0.484506, 0.483979, and 0.481773, respectively. These values are approximately the same since loss values directly express the coverage of models.

*Table 2: Complexity of estimators*

| Estimator | Computational complexity | Learning values | Training time (seconds) | Running time (seconds / sample) |
|---|---|---|---|---|
| DetNet: $K = 4$ | $\mathcal{O}(N_t N_r^2)$ | 263,688 | 46,972.63 | $1.34943 * 10^{-3}$ |
| ISDNN: $K = 4$ | $\mathcal{O}(N_t N_r^2)$ | 132,104 | 9,416.92 | $6.23742 * 10^{-5}$ |
| ISDNN: $K = 5$ | $\mathcal{O}(N_t N_r^2)$ | 165,130 | 11,134.20 | $1.33405 * 10^{-4}$ |
| S-ISDNN: $K = 4$ | $\mathcal{O}(N_t N_r^2)$ | 132,104 | 6,088.08 | $9.53519 * 10^{-5}$ |
| S-ISDNN: $K = 5$ | $\mathcal{O}(N_t N_r^2)$ | 165,130 | 8,157.22 | $1.40340 * 10^{-4}$ |

Table 2 shows the cost of DetNet-based CE, ISDNN, and S-ISDNN, in terms of computational complexity, number of learning values, training time, and running time. All considered estimators share the same computational complexity measured by Big-O notation at $\mathcal{O}(N_t N_r^2)$. The computational complexities of traditional estimators, i.e., LS and MMSE (Minimum Mean Square Estimation), are $\mathcal{O}(N_t N_r^3)$. Since they replaced this inversion of the LS estimator by learning values, so its complexity is $\mathcal{O}(N_t N_r^2)$. The number of DetNet's learning values is most significant at 263,688 parameters because it requires more linear operators than that of ISDNN. Then, ISDNN and S-ISDNN with 4 layers require 132,104 learning parameters, whereas that number if $K = 5$ is 165,130. It is essential to recognize that the number of learning values rapidly increases with $N_r$ since linear operator shape only depends on $N_r$ as shown in Table 1. More learning parameters lead to longer training time. The training time of ISDNN and S-ISDNN with $K = 4, 5$ are 9416.92, 11134.20, 6088.08, and 8157.22 seconds, respectively. The interesting point here is that although the NMSE is lower, the proposed S-ISDNN architecture gives significantly faster training time, about 70% that of ISDNN with the same number of layers. On the contrary, DetNet lasts around 4 times and over 7 times training time compared to ISDNN and S-ISDNN, respectively. This issue is not only due to a large number of learning values but also because the DetNet structure does not support batch size configuration. However, offline training time is not the most important factor for an estimator because this can be done on high-end computers. We then consider the running time of training models, which is the needed time for a session of estimating channel. This value is proportional to the number of parameters in trained models. Thus, DetNet is still the model with the slowest running speed at $1.34943 * 10^{-3}$ seconds per sample. Regarding proposed ISDNN, the ISDNN with 4 layers and S-ISDNN with 5 layers are fastest and slowest models at $6.23742 * 10^{-5}$ and $1.40340 * 10^{-4}$ seconds per sample. These results are vital in 5G and beyond due to the extremely large increase in sample rates of these networks.

Figure 3 depicts NMSEs of ISDNN, S-ISDNN, and DetNet-based CE at different SNR levels. The NMSE values of all estimators range from -2.73 dB to -2.3 dB. Similar to training results, the proposed ISDNN is the best model and outperforms DetNet. Learning-based estimators are beneficial as they are less influenced by SNR fluctuations. This property is particularly important in 5G due to the significant signal attenuation in the environment.

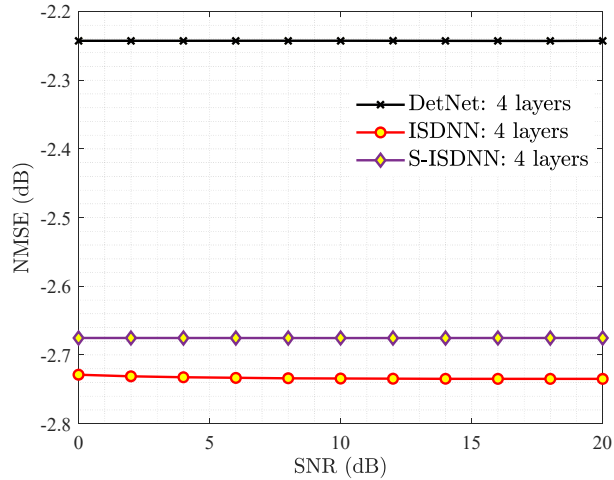

*Figure 3: NMSEs of ISDNN and DetNet*

## 5. CONCLUSION

In this paper, we introduced a DNN, termed ISDNN, designed for CE within massive MIMO systems. The ISDNN operates as a single-step CE network, eliminating the necessity for inverse operators. Simulation results demonstrate that ISDNN can converge during the learning process, achieving significantly high accuracy compared to the DetNet structure.

Additionally, thanks to side information-aided in 5G, we proposed the S-ISDNN network tailored for structured channel models. In addition, we also consider improving the performance of the ISDNN network by combining it with advanced data-driven deep learning network models, e.g., CNN, ResNet, and so on, in the future.

**ACKNOWLEDGMENT**

This work has been supported by VNU University of Engineering and Technology under project number CN21.04.

**REFERENCES**

[1]. Willhammar, S., Flordelis, J., Van Der Perre, L., and Tufvesson, F, 2020. *Channel hardening in massive mimo: Model parameters and experimental assessment*. IEEE Open Journal of the Communications Society, 1, 501–512.

[2]. Thakur, R. and Saxena, V. N., 2023. *A survey on learning-based channel estimation methods used for 5g massive mimo system*. 14th International Conference on Computing Communication and Networking Technologies (ICCCNT), July 2023, 1–5.

[3]. Dong, P., Zhang, H., Li, G. Y., Gaspar, I. S., and NaderiAlizadeh, N., 2019. *Deep cnn-based channel estimation for mmwave massive mimo systems*. IEEE Journal of Selected Topics in Signal Processing, 13, 989–1000.

[4]. Gao, J., Zhong, C., Li, G. Y., and Zhang, Z., 2022. *Deep learning-based channel estimation for massive mimo with hybrid transceivers*. IEEE Transactions on Wireless Communications, 21, 5162–5174.

[5]. Balevi, E., Doshi, A., and Andrews, J. G., 2020. *Massive mimo channel estimation with an untrained deep neural network*. IEEE Transactions on Wireless Communications, 19, 2079–2090.

[6]. Samuel, N., Diskin, T., and Wiesel, A., 2019. *Learning to detect*. IEEE Transactions on Signal Processing, 67, 2554–2564.

[7]. Mandloi, M. and Bhatia, V., 2017. *Low-complexity near-optimal iterative sequential detection for uplink massive mimo systems*. IEEE Communications Letters, 21, 568–571.

[8]. Hershey, J. R., Roux, J. L., and Weninger, F., 2014. *Deep unfolding: Model-based inspiration of novel deep architectures*. arXiv:1409.2574, 1–27.

[9]. Son, D. H. and Thuy Quynh, T. T. (2023) Impact analysis of antenna array geometry on performance of semi-blind structured channel estimation for massive MIMO-OFDM systems. IEEE Statistical Signal Processing Workshop (SSP), Hanoi, Vietnam, July 2023.

[10]. Ladaycia, A., Mokraoui, A., Abed-Meraim, K., and Belouchrani, A., 2017. *Performance bounds analysis for semi-blind channel estimation in mimo-ofdm communications systems*. IEEE Transactions on Wireless Communications, 16, 5925–5938.

[11]. Liu, L. and Yu, W., 2018. *Massive connectivity with massive mimo—part i: Device activity detection and channel estimation*. IEEE Transactions on Signal Processing, 66, 2933–2946.

[12]. Hammad, M. M. and Yahia, M. M., 2023. *Mathematics for Machine Learning and Data Science: Optimization with Mathematica Applications*. arXiv 2302.05964.

[13]. Liao, J., Zhao, J., Gao, F., and Li, G. Y., 2020. *A model-driven deep learning method for massive mimo detection*. IEEE Communications Letters, 24, 1724–1728.

[14]. Kingma, D. P. and Ba, J., 2015. *Adam: A method for stochastic optimization*. 3rd International Conference on Learning Representations, San Diego, CA, USA, pp. 1–15.

[15]. Swindlehurst, A. L., Zhou, G., Liu, R., Pan, C., and Li, M., 2022. *Channel estimation with reconfigurable intelligent surfaces—a general framework*. Proceedings of the IEEE, 110, 1312–1338.

**THÔNG TIN TÁC GIẢ**

**Đỗ Hải Sơn[1], Vũ Tùng Lâm[2], Trần Thị Thúy Quỳnh[2, *]**

[1]Viện Công nghệ Thông tin – Đại học Quốc gia Hà Nội

[2]Trường Đại học Công nghệ – Đại học Quốc gia Hà Nội